\documentclass[printer]{aa}
\usepackage{natbib}
\bibpunct{(}{)}{;}{a}{}{,} 
\usepackage{graphicx}
\usepackage{tabularx}

\begin{document}

\title{Extending the Canada-France brown Dwarfs Survey to the near-infrared: \\
first ultracool brown dwarfs from CFBDSIR
\thanks{Based on observations obtained with WIRCam, a joint project of
  CFHT, Taiwan, Korea, Canada, France, and the Canada-France-Hawaii
  Telescope (CFHT) which is operated by the National Research Council
  (NRC) of Canada, the Institute National des Sciences de l'Univers of
  the Centre National de la Recherche Scientifique of France, and the
  University of Hawaii.
Based on observations obtained with MegaPrime/MegaCam, a joint
  project of CFHT and CEA/DAPNIA, at the Canada-France-Hawaii
  Telescope (CFHT) which is operated by the National Research Council
  (NRC) of Canada, the Institut National des Sciences de l'Univers of
  the Centre National de la Recherche Scientifique (CNRS) of France,
  and the University of Hawaii. This work is based in part on data
  products produced at TERAPIX and the Canadian Astronomy Data Centre
  as part of the Canada-France-Hawaii Telescope Legacy Survey, a
  collaborative project of NRC and CNRS. 
  Based on observations made with the
ESO New Technology Telescope at the La Silla Observatory under
programs ID 082.C-0506(A) and 083.C-0797(A) with SOFI at NTT and ESO
VLT Director Discretionary Time program 282.C-5075 with ISAAC.
}} 
 

\author{
  P. Delorme \inst{1,4}
  \and L. Albert \inst{2}
  \and T. Forveille \inst{4}
  \and E. Artigau \inst{3,9} 
  \and X. Delfosse \inst{4}
  \and C. Reyl\'e \inst{5}
  \and C. J. Willott \inst{6}
  \and E. Bertin \inst{7}
  \and S. M. Wilkins \inst{10}
  \and F. Allard \inst{8}
  \and D. Arzoumanian \inst{11}
}

\offprints{P. Delorme, \email{pd10@st-andrews.ac.uk}}

\institute{School of Physics \& Astronomy,University of St Andrews
North Haugh,St Andrews KY16 9SS, Scotland
    \and Canada-France-Hawaii Telescope Corporation, 65-1238 Mamalahoa 
   Highway, Kamuela, HI96743, USA
    \and D\'epartement de physique and Observatoire du Mont M\'egantic,
  Universit\'e de Montr\'eal, C.P. 6128, Succursale Centre-Ville,
  Montr\'eal, QC H3C 3J7, Canada
\and Universit\'e Joseph Fourier - Grenoble 1, Centre national de la
recherche scientifique, Laboratoire d'Astrophysique de Grenoble
(LAOG), UMR 5571, 38041 Grenoble Cedex 09, France 
  \and Observatoire de Besan\c{c}on, Institut Utinam, UMR CNRS 6213, 
   BP 1615, 25010 Besan\c{c}on Cedex, France 
  \and  Herzberg Institute of Astrophysics, National Research Council, 5071 West Saanich Rd, Victoria, BC V9E 2E7, Canada
  \and Institut d'Astrophysique de Paris-CNRS, 98bis Boulevard Arago, 
   F-75014, Paris, France
  \and C.R.A.L. (UMR 5574 CNRS), Ecole Normale Sup\'erieure, 69364 Lyon
  Cedex 07, France
\and Gemini Observatory Southern Operations Center c/o AURA, Casilla 603
   La Serena, Chile
\and Oxford Astrophysics, Department of Physics, Denys Wilkinson Building
Keble Road Oxford, OX1 3RH, United Kingdom.
\and  CEA-Saclay, IRFU, SAp, 91191, Gif-sur-Yvette, France
}

\abstract{}
{
We present the first results of the ongoing Canada-France Brown Dwarfs
  Survey-InfraRed, hereafter CFBDSIR, a near infrared extension to the
  optical wide-field survey CFBDS. Our final objectives are to constrain
  ultracool atmosphere physics by 
finding a statistically significant sample of objects cooler than 650K
and  to explore the ultracool brown dwarf mass function building
on a well-defined sample of such objects.}
{We identify candidates in CFHT/WIRCam $J$ and CFHT/MegaCam $z'$
  images using optimised psf-fitting, and follow them up 
with pointed, near-infrared imaging with SOFI at the NTT. We finally
obtain 
low-resolution spectroscopy of the coolest candidates to characterise
their atmospheric physics.} 
{ We have so far analysed and followed up all candidates on the first
   66 square degrees of the 335 square degree survey. 
We identified 55 T-dwarfs candidates with $z'-J>3.5$ and have
confirmed six of them as T-dwarfs, including 3 that are strong
later-than-T8 candidates, based on
their far-red  and NIR colours. We also present here the NIR spectra
of one of these 
 ultracool dwarfs, CFBDSIR1458+1013, which confirms it as one of the coolest
 brown dwarf known, possibly in the 550-600K temperature range. }
{From the completed survey we expect to discover 10 to 15 dwarfs 
later than T8, more than doubling the known number of such objects.
This will enable detailed studies of their extreme atmospheric 
properties and provide a stronger statistical basis for studies of their
luminosity function.
}

\date{}


\titlerunning{CFBDSIR: extending the CFBD Survey to the near-infrared}

\maketitle

\section{Introduction}

The significant improvement in detector technology, data storage, 
 and analysis facilities in
 the past decade has made it possible to carry 
out wide-field surveys covering a large fraction of the sky instead of
targeting specific sources. The wealth of data from these surveys
necessitate a complex dedicated computer analysis to single out
relevant scientific information. These surveys, such as DENIS \citep{Epchtein.1997}, SDSS
 \citep{York.2000}, 2MASS \citep{Skrutskie.2006}, UKIDSS
 \citep{Lawrence.2007}, and CFBDS \citep{Delorme.2008b} contain
 hundreds of millions of 
 astrophysical sources and led to many advances in various fields,
 notably to identify extremely rare objects and build robust
 statistical studies. 

The survey we are presenting here, the Canada-France Brown Dwarfs
Survey-InfraRed aims at finding ultracool brown dwarfs (T$_{eff}<$
650K) of
which only 6 are currently published by
\citet{Warren.2007,Delorme.2008a,Burningham.2008,Burningham.2009,Burningham.2010sub}. 
 \citet{Lucas.2010sub} have very recently identified a probably even 
cooler object. These
rare objects are in many ways the intermediate ``missing link" between the cold
atmospheres of the Solar System's giant planets and
cool stellar atmospheres. The physics and chemistry of their
atmospheres, dominated by broad molecular absorption bands, are very
planetary-like \citep[see][for instance]{Kulkarni.1997} and the cool
brown dwarfs spectra are the key to constraining planetary and stellar
atmosphere models.

 Nowadays, the Teff$<$700K atmosphere temperature regime is troublesome
 for modellers.   
A few ultracool late T brown dwarfs have now been discovered  with
effective temperatures below 650K. These 
discoveries step into unexplored territory and a new generation of 
models is emerging. It is facing several difficulties. 1)
Out-of-equilibrium chemistry plays an important role, resulting for
instance  
in NH$_3$ being less prevalent than expected \citep{Cushing.2006}. 2) Fine details 
of convection control both the L/T transition and the dredge up of 
hot chemical species in late T atmospheres. 3) Water cloud formation 
and dust nucleation play important roles. 4) Line opacities of several
molecules, in particular NH$_3$ and, to a lesser extent, CH$_4$ are
unknown and cause important spectroscopic feature mismatches.  As a
good example of the need for refined models, \citet{Burningham.2009} 
have recently determined that a T8.5 dwarf companion to an M star was
actually $\sim$15\% cooler than model fitting would have
predicted and \citet{Dupuy.2009} and  \citet{Liu.2008} reached similar
conclusions when studying brown dwarf binaries with dynamical masses. 

Under those circumstances, observations are key to the development of 
the models. Only five brown dwarfs with temperatures below 650K (T8.5) are 
currently known from recent discoveries by UKIDSS and CFBDS, and this small 
number prevents discerning general trends from individual 
peculiarities. \citet{Kirkpatrick.1999} and \citet{Burgasser.2002}
could rely on samples of 20-25 objects to respectively define the 
L and T spectral types.
 In this article we present the CFBDSIR, a near infrared (hereafter
 NIR) extension to
 the CFBDS that will provide a WIRCAM \citep{Puget.2004} $J$-band coverage down to $J_{vega}=20.0$
 for 10$\sigma$ detections atop 335 square
 degrees of CFBDS MegaCam \citep{Boulade.2003proc} $z'$-band imaging with a 5$\sigma$ detection limit of
 $z'_{AB}$=23.25-24.05. All optical magnitudes presented in this
 article are AB magnitudes, 
while all NIR magnitudes are Vega magnitudes.
When the CFBDSIR is complete, we hope to achieve a threefold increase
 in sample size to 15-20 characterised ultracool brown dwarfs and
 possibly find a few substellar objects significantly cooler than 500K of which
 none is known yet outside the solar system. This will define
 general trends and dispersions around  
them, permitting the study of ultracool dwarfs not only as individual
interesting objects, but as a population. This will help define the
T/Y spectral transition that is expected to occur in this temperature
range. In section 1 we present the rationale of this new wide-field
survey and the observations at its core. In section 2 we describe the
data reduction and the data analysis methods we used to identify
ultracool brown dwarfs candidates. Finally, we present the spectra and
the photometry of
the first ultracool brown dwarf identified with CFBDSIR in section 3.

\section{Observations}
  \label{observations}  
\subsection{Far-red and near infrared photometric properties of brown dwarfs}
 Field brown dwarfs are cool objects, with a temperature range that
currently extends from $\sim$2500K (early L) to $\sim 525$K
\citep[latest objects identified, e.g.][]{Leggett.2009}. 
Even cooler, and still to be found, brown dwarfs should close the temperature gap between 
late type T-dwarfs and solar system Jovian planets($\sim$110K). These
objects do exist \citep[see][for instance]{Burgasser.2009proc} because
low-mass brown dwarfs already observed in young clusters must cool
down to this temperature range when they age. 
Brown dwarfs spectra differ significantly from a black body, and have considerable
structure from deep absorption lines and bands. The spectral energy
distribution of 500K brown dwarfs peaks in the in the $J$ band in the
near infrared 
(even colder objects should emit more energy in the
mid-infrared), and they are therefore most easily detected
in that wavelength range. Their NIR $JHK$ colours are, however,
 not distinctive at a modest signal-to-noise ratio \citep[see][for instance]{Metchev.2008}, and brown dwarfs are more
easily recognised by including at least one photometric band
blueward of 1~$\mu$m. At those wavelengths the steep slope 
of their spectra stands out, and they have very distinctively
red $i'-z'$ and $z'-J$ colours. The CFBD Survey \citep{Delorme.2008b}
took advantage of their distinctive $i'-z'$ colours to identify
hundreds of L and T-dwarfs.  However, the  reddest
and coolest  brown dwarfs have extremely red $z'-J$ colours and are
much easier to detect in the NIR than on original CFBDS far-red
images. A WIRCam $J$-band coverage of CFBDS 
fields is a straightforward 
and efficient way to cumulate both the very effective selection
criteria using far-red colour information and the NIR detection sensitivity to
ultracool brown dwarfs. Since the overlap of 
CFBDS with existing NIR surveys with a relevant depth \citep[such as
  UKIDSS,][]{Lawrence.2007} is marginal, it was impossible to rely on
archive data.

 The resulting survey, CFBDSIR, identifies astrophysical sources on 
$J$-band WIRCam \citep{Puget.2004} images and selects ultracool brown dwarfs candidates 
depending upon their $z'-J$ colours, using $z'$ magnitudes from CFBDS
MegaCam \citep{Boulade.2003proc} images. 
The $z'-J$ colour has a wide range for brown dwarfs, varying
from 2.6  for mid-L types to  
over 4.5 for late-T types (Fig.~\ref{zJspT}). It therefore
provides (at least at a good signal-to-noise  ratio) a good selection criteria to
identify ultracool brown dwarfs. They usually have $z'-J>$3.8, as
confirmed by atmosphere models such as BT-Settl \citep{Allard.2007}, synthetic
MegaCam/WIRCam colours derived from known objects
\citep{Delorme.2008b}  and direct 
observational data \citep{Delorme.2008a, Burningham.2008,
  Burningham.2009}. Models 
suggest that cooler objects, not yet discovered (T$<$500K), are even redder.
 Figure ~\ref{zJspT} however shows that the colour-spectral type relation 
 for late T-dwarfs has a wide spread, meaning that
 this $z'-J$ colour range also contains some warmer mid-T 
 dwarfs. Follow-up photometry of candidates in $H$ and $K_{\rm s}$
 band easily distinguishes the relatively blue very late T from the redder
 mid-T-dwarfs \citep[see][for instance]{Lodieu.2007}. Only very rare
 astrophysical objects share the same $z'JHK$ colour  
range:  very high-redshift starburst-galaxies, extremely red Balmer break
galaxies, and atypical broad absorption-lines quasars at 
moderately high redshift
($z\geq 2$), such as those described in \citet{Hall.2002}. However,
those objects can be differenciated from brown dwarfs using only
photometry: they are not as red in   $i'-z'$ as ultracool brown dwarfs
and will appear in the $i'$ CFBDS images whereas brown dwarfs will
not.  Quasars with redshift z$>$7 cannot be distinguished from 
cool brown dwarfs using the CFBDSIR $i',z',J$ and optical
photometry, but we would be lucky to find just one of these
extremely interesting objects. None has been 
discovered yet, and extrapolation of the z$=$6 luminosity
function \citep{Willott.2010} suggests that wider or/and deeper 
surveys will be needed to find one. 
Other possible contaminants are detector artefacts
and variable objects whose luminosity changed between $z'$ and $J$
exposures.

      \subsection{The survey} 
{\bf WIRCam NIR data.} 
This article focuses on the results obtained from a 66 square degrees
pilot subset of CFBDSIR whose final coverage is expected to be 335
square degree.  
CFBDSIR WIRCam imaging goes to a depth of
$J_{vega}=20.0$ for a point-source detection limit of 10$\sigma$, which
ensures
accurate photometry and rejects most spurious detections. Images were
acquired
at CFHT in QSO mode with seeing varying from 0.6\arcsec to 1.0\arcsec
during
semesters 2006B, 2007A, and 2009B. These WIRCam images overlap the
existing
$z'$-band images of CFBDS; i.e.,  each $\sim$ 1 square degree MegaCam field
is
covered by nine 21' by 21' WIRCam $J$-band 45-second exposures. For the
sake of
efficiency, no dithering is done. Instead, the 9 adjacent fields observed
in
sequence are used for sky construction. Each 9-field patch is actually
observed
twice, generally one night apart, in order to identify and remove any
moving
solar system object, for a total integration time of 90 seconds. The
second sequence is observed with few pixels offset to average the fixed noise
pattern
seen on the WIRCam detectors. 
 The  10$\sigma$ limiting depth of $J_{vega}=20.0$ ensures  that all
confirmed ultracool brown dwarf candidates can be observed with
low-resolution spectrographs on 8-meter class telescopes.  \\

\begin{figure}
\includegraphics[width=8cm]{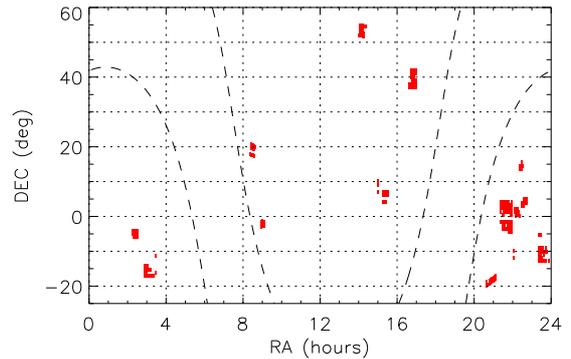}
\caption{Map of observed CFBDSIR fields as of March 2010. Dashed lines
  are +/-20 degrees of galactic latitude.
\label{cfbdsir_map}}
\end{figure}

{\bf Optical data.}
The optical data consist of the $z'$-band images used for the CFBD
Survey. This survey is made of public and P.I. data and associated 4
sub-surveys of different area shallowness, namely the Canada-France-Hawaii
Telescope Legacy Survey-Very Wide,CFHTLS-Wide, CFHTLS-Deep and
Red-sequence Cluster Survey-2 (RCS2), described in detail by
\citet{Delorme.2008b}. The CFBDSIR only targets the 3 shallower
components, as shown in Table \ref{maglim}. This table compares 
 the limiting magnitude, the maximum distances of detection for 
 mid-L, mid T and the latest T-dwarfs, together with
 the areas covered by each survey. The similar table 1 in
 \citet{Delorme.2008b} incorrectly uses the maximal magnitude limit
 for the 
 CFHTLS-WIDE, instead of the average one. The value of 23.3 stated in
 Table  \ref{maglim} here corrects this error. The 335 square degrees CFBDSIR
 probes about 2.8 times more volume for
 late T-dwarfs than the $\sim$900 square degree CFBDS,  and even more for
 redder (and thus likely cooler) objects.  Figure
   \ref{cfbdsir_map} shows the sky coverage of CFBDSIR as of March 
  2010. The optical data is public and available on the CFHT archive, 
  while the NIR data becomes public after the standard one-year
  CFHT proprietary period.\\

\begin{table*}
\begin{center}
\caption{Comparison of CFBDS optical surveys with CFBDSIR.\label{maglim}}
\begin{tabularx}{\textwidth}{|X|X|X|X|X|X|} \hline
 Survey Name&  $z'_{AB}$ $10\sigma$ ($5\sigma$) detection limit& mid-L maximal
 detection (pc) & mid-T maximal detection (pc) & T9 maximal detection
 (pc) & Final coverage (sq deg)\\ \hline \hline
 CFBDS:RCS-2            & 22.5 (23.25) &  185 & 80  &16 & 600  \\\hline
 CFBDS:CFHTLS-Very Wide & 22.8 (23.55) & 215 & 90 & 18  & 150  \\\hline
 CFBDS:CFHTLS-Wide      & 23.3 (24.05) & 270 & 115 & 22 & 170  \\\hline\hline
 Survey Name&  $J_{vega}$ $10\sigma$ detection limit & mid-L maximal
 detection (pc) & mid-T maximal detection (pc)& T9 maximal detection (pc) & Final coverage(sq deg)\\ \hline \hline
 CFBDSIR  & 20.0  & 235 & 145  & 35 & 335 \\\hline
\end{tabularx}
\begin{list}{}{}
\item CFBDS  initial detections are made in $z'$ and only better than
$10\sigma$ detections are used.
 CFBDSIR makes the initial detection in $J$, using only better than
$10\sigma$ detections in $J$, and is only limited in $z'$ by no
 detections. We define any $z'$-band object with a signal below the
 $5\sigma$ threshold as a no detection or ''z'-dropout".
\end{list}
\end{center}
\end{table*}

{\bf Survey rationale.}
Ultracool brown dwarfs have $z'_{AB}-J_{vega}>3.8$ for the MegaCam 
and WIRCam photometric systems, with some already observed 
at $z'-J=4.5$.  
Given the relative depth of the optical and NIR exposures, most of the
T-dwarfs earlier than T7 detected in the $J$-band image will have a
$z'$-band counterpart. The objects detected at $J$ and not at
$z'$ (hereafter $z$-dropouts) are thus good candidates for some of the
coolest T-dwarfs known ($>$T7), and possibly even cooler Y dwarfs.

 Since our most interesting candidates are likely to only be 
 detected in $J$-band, we need to carefully eliminate most sources of
 contamination in this band. The main contaminants are detectors artefacts,
 variable objects and asteroids. As described in
 detail in section \ref{data_analysis}, our point spread function
 (hereafter PSF)
 analysis removes most artefacts, while all asteroids are eliminated
 because they move between the two exposures taken at each
 pointing. Supernovae and other strongly variable objects have to 
 be rejected
 during the follow-up, but their numbers are kept low thanks to the
 relative shallowness of the survey.  Since we select
   all PSF-looking $z'$-band dropouts, CFBDSIR also very efficiently
   identifies high-proper motion objects: if the proper motion is high
   enough that a $J$-band source cannot be cross-matched with its
   $z'$-band counterpart, the automatic pipeline keeps the
   candidate in the selection as a $z'$-band
   dropout. Visual examination of all selected candidates subsequently
   identifies those high proper motion sources easily.\\

{\bf NIR Follow-up.}
The NIR follow-up of CFBDSIR candidates removes the remaining
candidates but also provides an efficient characterisation of the
confirmed candidates. Ultracool brown
dwarfs can be singled out by their blue $J-H$ and $H-K$ colours owing to
stronger H$_2$O, CH$_4$ and probably NH$_3$ absorption than earlier T
dwarfs. The CFBDSIR follow-up of the first  66 square degrees from
CFBDSIR required about 3.5 nights on SOFI
\citep{Moorwood.1998} at
NTT.
 We were able to observe all 55 T-dwarfs candidates
identified and to confirm   6 of
them as T-dwarfs, including 3 ultracool brown dwarfs (later than T7
dwarfs and possible Y dwarfs) candidates.
 The rationale of the NIR follow-up is the following.
\begin{itemize}
\item A short (5 to 10 minute dithered image, with 30-s individual
  exposures), $J$-band exposure confirms the candidate and eliminates
  any remaining contaminants from variable sources, such as supernovae
  or detector artefacts. For confirmed
  candidates, this image also provides good signal-to-noise (from 20
  to about 100, depending on the target magnitude)  $J$-band
  photometry, better than the photometry on the detection image.
\item All candidates confirmed with this SOFI $J$ photometry are imaged in $H$-band to characterise
  them as ultracool brown dwarfs or earlier T-dwarfs. A 10-20 minute
  dithered sequence, with 20-s individual exposures, usually achieve good
  signal-to-noise photometry, with an accuracy better than 5\%.
\item All confirmed candidates with measured $J-H<$0.1 are potential
  ultra-cool brown dwarfs that can be further characterised by acquiring
  $K_{\rm s}$-band photometry, if enough observing time is available.  A 15-30 minute
  dithered sequence, with 15-s exposures typically achieves good
  signal-to-noise photometry in $K_{\rm s}$.
\end{itemize}

 It should be noted that this very efficient confirmation and
 characterisation photometric follow-up is made possible  by
 visitor-mode observation  combined with the automatic 
 image reduction by SOFI pipeline \textit{Gasgano}. A simple 
 photometric analysis can be carried out by the observer only minutes after the
 exposures allowing $J$ and possibly $H$ and $K_{\rm s}$ photometry
to be derived on-the-fly. It is then possible to choose the follow-up
strategy most suited 
 to each individual candidate. The final astrometric and photometric
 calibration is done using 2MASS \citep{Skrutskie.2006} stars in the
 field as references.


\section{Survey analysis}
Ultracool brown dwarfs are extremely red objects 25 to 60 times (3.5 to
4.5 magnitudes) brighter in $J$-band that in $z'$ band. This means that
even if our $z'$ images are about 3 to 4 magnitudes deeper than the
detection $J$-band images, many of our candidates with strong $J$ band
detection will not appear at all in the $z'$ images. However, a non
detection in $z'$ band is a very strong sign that the candidate is a
very red astrophysical source with $z'-J>3.5$, exactly  what we are
looking for. The other alternative is that these
sources are spurious, such as unrecognised artefacts (remaining
detector cross-talk, 
unflagged hot pixel, optical ghost, etc.) on the $J$ image. Since
ultracool  brown dwarfs
are extremely rare ( our preliminary estimate is $\sim$ 1 per 25
square degree, so fewer than 1 per $500~000$ sources down to $J$=20.0),
false detection rates of 
even 1 per 10$^4$ real sources would greatly outnumber true brown dwarfs. We therefore
need a very 
robust analysis of the discovery images to tie down the number of
artefacts before we finally weed them out during the photometric
follow-up of the candidates. 

\subsection{Image reduction}
 The WIRCam observations targeted 335 existing MegaCam
fields from CFBDS, that have already been observed in $i'$ and
$z'$. The reduction of MegaCam data is described by
\citet{Delorme.2008b}. 

Images were run through the `I`iwi Interpretor of the WIRCam Images
(`I`iwi -
{\tiny www.cfht.hawaii.edu/Instruments/Imaging/WIRCam})
at
CFHT, which does non-linearity correction, dark subtraction, flat-fielding,
sky
subtraction, bad pixel masking, photometric calibration, and rough
astrometric
calibration.  Cross-talk subtraction was also performed on the 06B and
07A data, affected by amplifier to amplifier cross-talk.

Each set of 2 times 9 WIRCam images
covering one $z'$ MegaCam image was then co-aligned, using the $z'$ image
used as the astrometric reference. This astrometric calibration was
performed with \textit{Scamp}
\citep{Bertin.2006}, and the images were stacked 
using  \textit{Swarp}\citep{Bertin.2002}. This produced $\sim$1
square degree WIRCam images aligned with each targeted MegaCam field.

 \subsection{Image analysis} \label{data_analysis}
Since our stacks are only 2 exposures deep, the final frame is the
average of both exposures. To easily reject moving solar
system objects from our detection catalogues, we also produced a ''minimum"
image of both exposures. This combined image contains
the lower of the pixel values of the two exposures, 
and is effectively devoid of all those objects that moved
by more than one FWHM between the exposures. We carried out the
analysis using the double-image mode of \textit{Sextractor} 
\citep{Bertin.1996}, with source detection on the ''minimum" 
image and photometric measurements
on the average image, using PSF fitting. The different analysis steps are described here: 
\begin{itemize}
\item Spatially variable PSF models of the images are built with
  PSFex \citep{Bertin.2010}, using single stars within the image 
  as prototypes..
\item The sources on the minimal image are identified using
  \textit{Sextractor}. Their astrometry and photometry  are derived 
  by PSF fitting. (The latter is
  discarded from thereon, but is included in the fit).
Since moving objects are at different positions in the two exposures
that were stacked to produce the minimal image,
they do not appear in  the resulting source list.
\item  A flux-only PSF fitting at the position of the sources identified on the
  minimal sum stacked image is then carried out on the average
  stacked image. This step produces the final WIRCam $J$-band catalogue.
\item The MegaCam $z'$-band catalogue of the corresponding field is then
  produced using a similar position and flux PSF fitting with \textit{Sextractor}.
\item The $J$-band and  $z'$-band catalogues are cross-matched
  so that each source in the $J$-band catalogue is associated with a
  $z'$ magnitude. In case there is no $z'$-band couterpart to a
  $J$-band detection, which is then a ``$z$-dropout", the source is given the 5
  sigma detection magnitude limit of the full $z'$ image as its $z'$
  magnitude and the
  corresponding $z'-J$ colour is then considered as a lower limit of
  its actual colour.

\end{itemize}
 
  \subsection{Filtering and candidate selection}
  The selection of ultracool brown dwarfs candidates in our catalogue
  aims at the greatest possible completeness (i.e. finding most of the 
  ultracool dwarfs actually detectable on the images) while keeping
  the number of contaminants down to an acceptable level. 
  Contaminants are particularly critical because the most promising
  candidates are the $z$-dropouts for which only a $J$ magnitude and a
  $z'-J$ lower limit on the colour is known. To keep the number of
  contaminants relatively low, we only selected sources with a
  signal-to-noise ratio above 10 in the $J$-band stack.

\begin{figure}
\includegraphics[scale=0.33,angle=0]{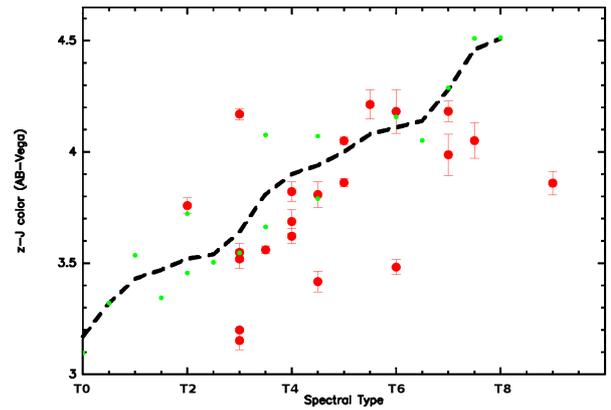} 
\caption{$z'-J$ colour spectral type relation. The green symbols represent
  synthetic MegaCam/WIRCam colours for known objects with
  publicly available spectra.
The black dashed line is the resulting averaged colour-spectral type
relation. Red points show the observational colours and spectral types
of CFBDS spectroscopically confirmed T-dwarfs.  
\label{zJspT}     }  
\end{figure}

\begin{figure}[!h]
\begin{center}
\includegraphics[scale=0.36,angle=0]{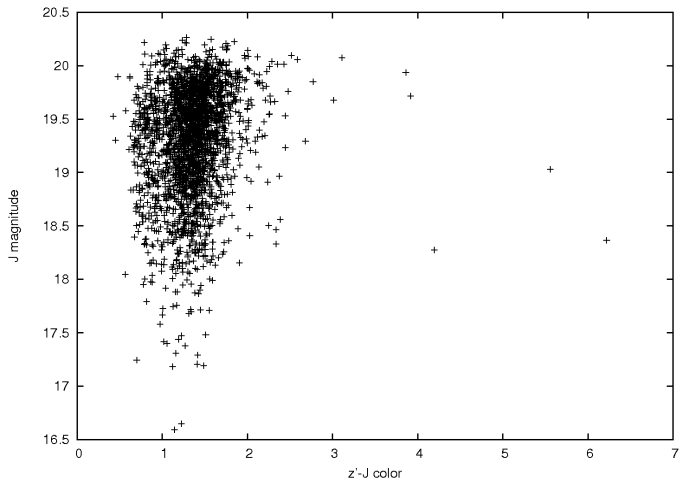}
\caption{Colour-magnitude diagram of all sources detected with a signal-to-noise
  greater than 10 in a 1 square-degree stacked WIRCam image from
  CFBDSIR. Since no other filtering is applied, these sources can be 
  stars, galaxies, or detector artefacts. In this specific instance,
  none of the red objects is a brown dwarf.
\label{colpop}}
\end{center}
\end{figure}

\begin{figure}[!h]
\begin{center}
\includegraphics[scale=0.6,angle=0]{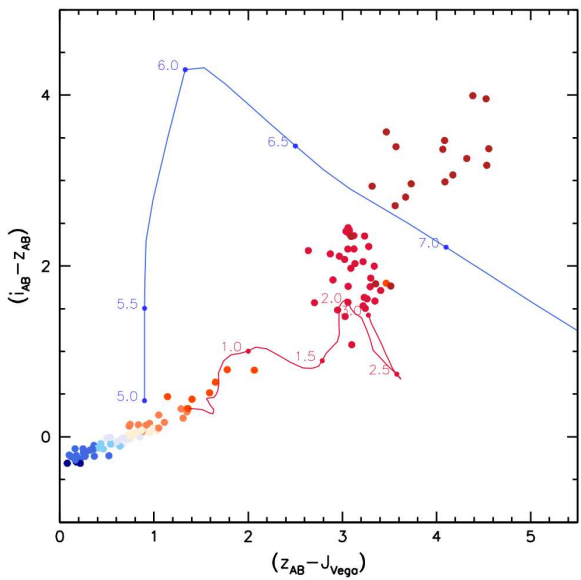}
\caption{Colour-colour diagram of known stars and brown dwarfs from spectral
  type O (deep-blue dots) to T (dark-red dots). The blue line shows the
  colours of high-redshift-starburst galaxies, while
  the red line refers to red Balmer-break galaxies. These lines follow
  a redshift evolution and specific redshift steps are indicated by
  numbers near the lines.  
  The colours are synthesised from spectra, and take into account the
  actual transmission and detector sensitivity of the MegaCam and
  WIRCam instruments used in CFBDSIR.
\label{wilkins}}
\end{center}
\end{figure}

  As visible in Fig. \ref{zJspT}, brown dwarfs later than T8
  populate the $z'-J>$3.8 colour range. This figure as well as Fig.
  \ref{wilkins} make use of spectra from
  http://www.jach.hawaii.edu/$\sim$skl/LTdata.html
  \citep{Martin.1999,Kirkpatrick.2000,Geballe.2001,Leggett.2002,Burgasser.2003,Knapp.2004,Golimowski.2004,Chiu.2006}. As
  described in \citet{Delorme.2008b} these colours are synthesised from
  the spectra using the filters, atmosphere, telescope and detector
  transmission and sensitivity of the instruments used in CFBDSIR.
  Given the large colour spread of
  late T-dwarfs, we set a very conservative colour selection
  threshold of $z'-J>$3.5, to select as many
  ultracool dwarfs as possible. Since the 3.5$<z'-J<$3.8 colour range  
  is mainly populated by warmer T-dwarfs and is still very far from any
  densely populated colour locus, as illustrated by Fig.
  \ref{colpop}, including it in our selection does not increase
  the number of contaminants significantly.

In addition to this colour selection, the main filtering criterion is 
based on the $\chi ^2$ from PSF fitting, to distinguish point sources
from galaxies and artefacts, following the same method as described in
\citet{Delorme.2008b}. Since brown dwarfs are point sources, this
morphological rejection of all non-point-source-like objects
effectively removes many non-stellar contaminants. 

This automated selection produces an average of 6 candidates per square 
degree. Each of these is visually examined to remove remaining 
contaminants. In addition to inspecting both the $z'$-band and the 
$J$-band images, we examine the $i'$-band image (available from the 
CFBDS), to reject some variable contaminants, as well as possible 
contamination by atypical broad absorption line quasars, none of 
which is known to be very red in both $z'-J$ and $i'-z'$ 
\citep[P.Hall, private
  communication and ][]{Hall.2002}. Since the 5$\sigma$ detection
limits of the CFBDS $i'$ images (between 24.5 and 25.5) are significantly
deeper than the $z'$ images, broad absorption line quasars with
$z'-J>3.5$ should be visible on the $i'$ image. We identified one
such object, CFBDSIR232451-045852, with $z'-J=$4.7 and $i'-z'$=0.9.

Very high-redshift (z$>$6), star-forming galaxies and evolved galaxies
with strong Balmer/4000\AA ~breaks at z$\sim$2, shown in Fig.
\ref{wilkins}, are other possible contaminants. While the surface
density of J$\sim$20  high-redshift, star-forming galaxies 
is unknown (the brightest of these objects discovered so far have
J$_{vega}\sim$25 \citep{Ouchi.2009}, five magnitudes fainter than our
detection limit), the expected exponential cut-off in the luminosity function
 suggests that the probability of
finding any such galaxy in CFBDSIR is extremely low. 
 Balmer/4000\AA ~break galaxies at z$\sim$2.5 do enter the fringe
of our selected colour range; however, their $i'-J$ colour is typically
below 5, ensuring they are detected in the $i'$-band CFBDS images and
are then eliminated from our candidate list.

After this final visual check, we end up with about 1 ultracool
brown dwarf candidate per square degree, which has to be confirmed
with pointed NIR observations.

\section{Results}
\subsection{Photometric confirmation of 3 ultracool brown dwarfs}
The 55 candidates found in a 66 square degree subset of
 the data were followed-up with photometry at the ESO New
Technology Telescope, using the SOFI  near infrared camera during
visitor mode runs 083.C-0797(A) and 082.C-0506(A), in
July 2009 and March 2009. These pointed NIR observations confirm
six T-dwarfs, of which 3 are robust ultracool brown dwarf
candidates. Those 
are likely to be either very late T-dwarfs ($>$T8) or cooler Y
dwarfs. The  
brighter 2 of the 3 earlier T candidates are re-identifications
of CFBDS-discovered brown dwarfs, spectroscopically confirmed as 
T3.5 and T5  (Albert et al., submitted.).

  The candidates identified in the remaining 269 square degrees have 
  not yet been followed-up, and they certainly include mostly contaminants.
  Extrapolating the results from the 66 square degree pilot subset 
  of CFBDSIR to the full survey, we expect to find $\sim$15 ultracool 
  brown dwarfs. Low-resolution  NIR spectroscopy of the confirmed 
  ultracool brown dwarfs will be needed to derive spectral types and
  characterise their cool atmosphere physics. \\

{\bf Photometric properties of ultra-red brown dwarf candidates}

\begin{figure}
\includegraphics[width=8cm]{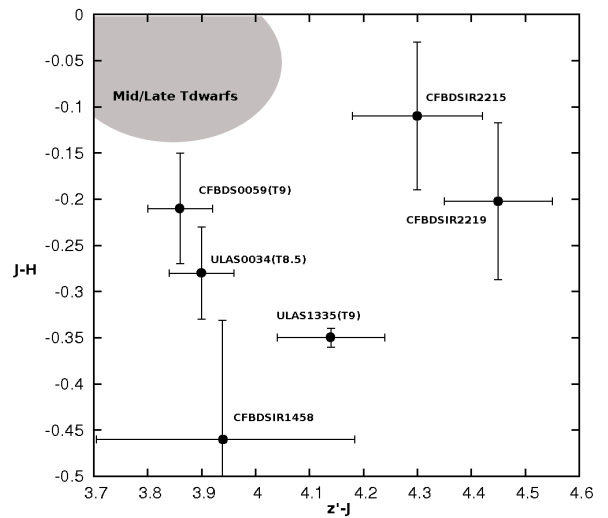}
\caption{Colour-colour diagram of some of the latest brown dwarfs
  known, $>$T8, later than spectrum used for up to date spectral
  classification scheme together with the 3 CFBDSIR ultracool dwarfs
  candidates. The regular T-dwarf approximate colour range is highlighted in the
  upper left corner. Note that given the strong dispersion in colours
  of T dwarfs, some mid/late T do spread out of this
  indicative colour range.
\label{zJJH}}
\end{figure}

After the March 2009 $J, H$, and $K_s$ follow-up observations, 
one candidate stood out as particularly interesting. We then
obtained additional WIRCam $J, H$, and $K_s$ photometry for this
object, CFBDSIR J145829+101343 (hereafter CFBDSIR1458). The 
resulting magnitudes are presented in Table~\ref{mag}. The SOFI
photometry in this table has additional uncertainty because its
calibration is bootstrapped from a small number of good 2MASS
reference stars in the narrow field of the SOFI NTT images, especially
so in the $H$ and  $Ks$ bands.

\begin{figure*}[h!]
\begin{tabular}{c|c}
 \includegraphics[scale=0.4]{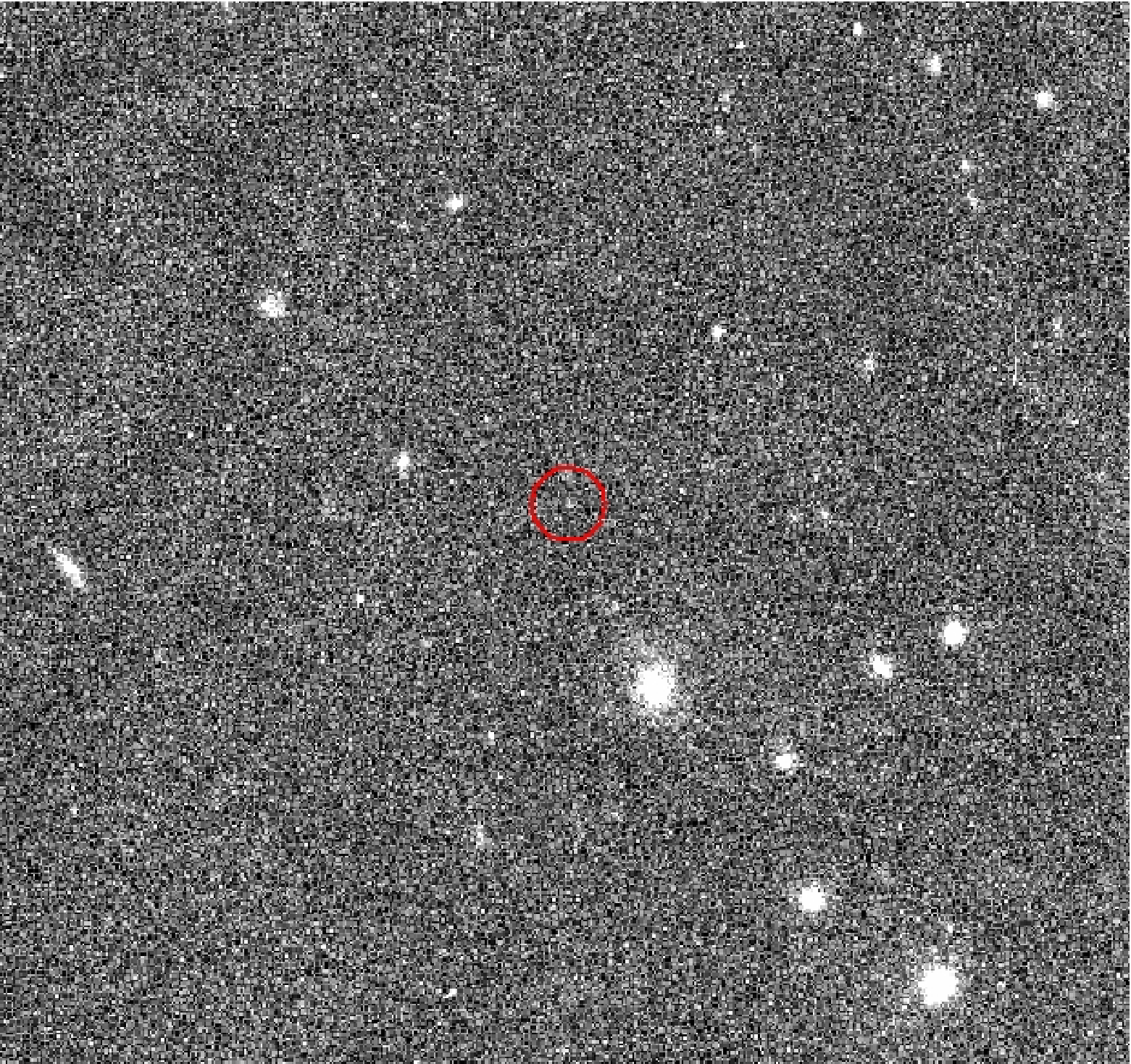}& 
 \includegraphics[scale=0.4]{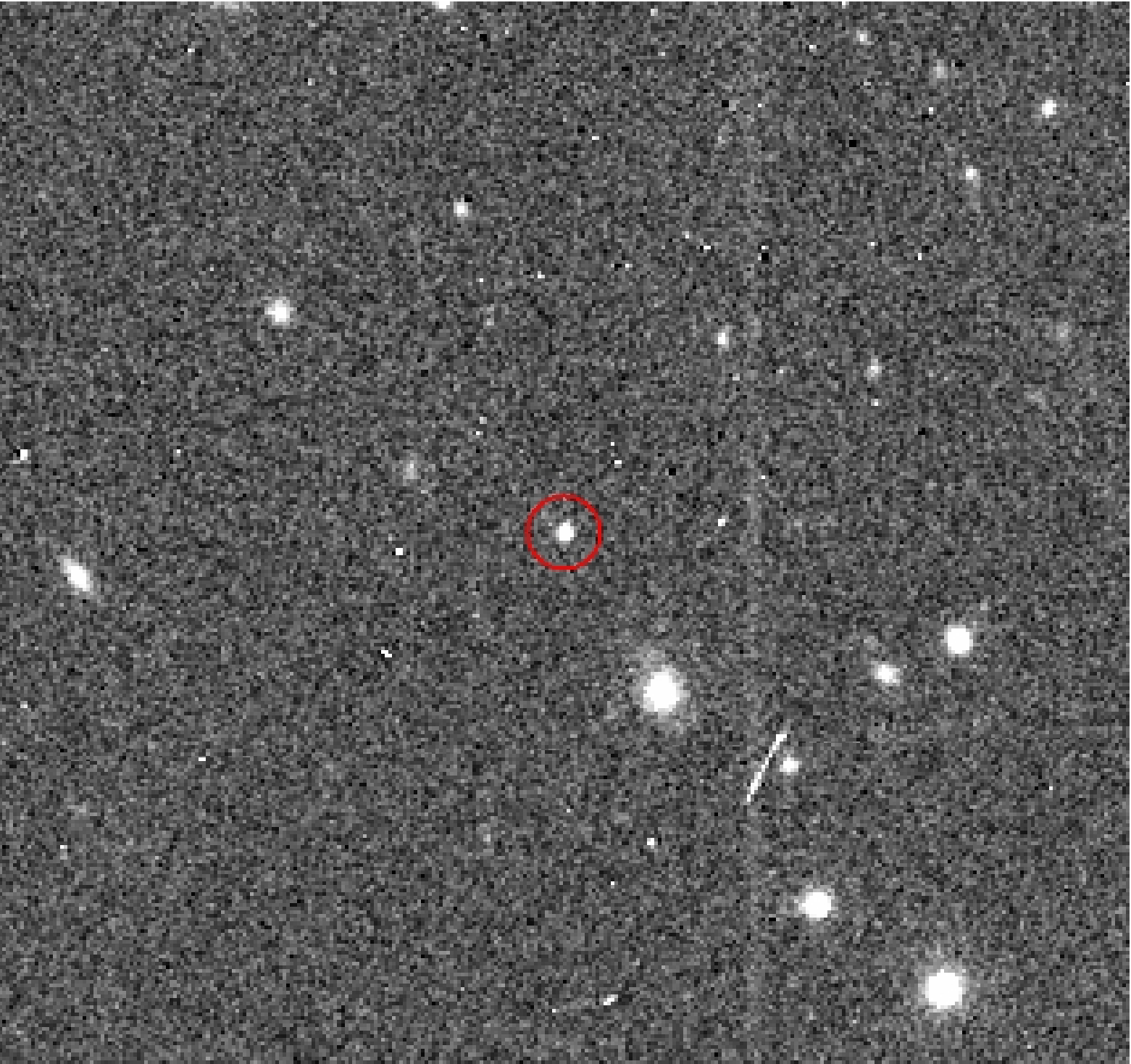}\\ \hline
 \includegraphics[scale=0.4]{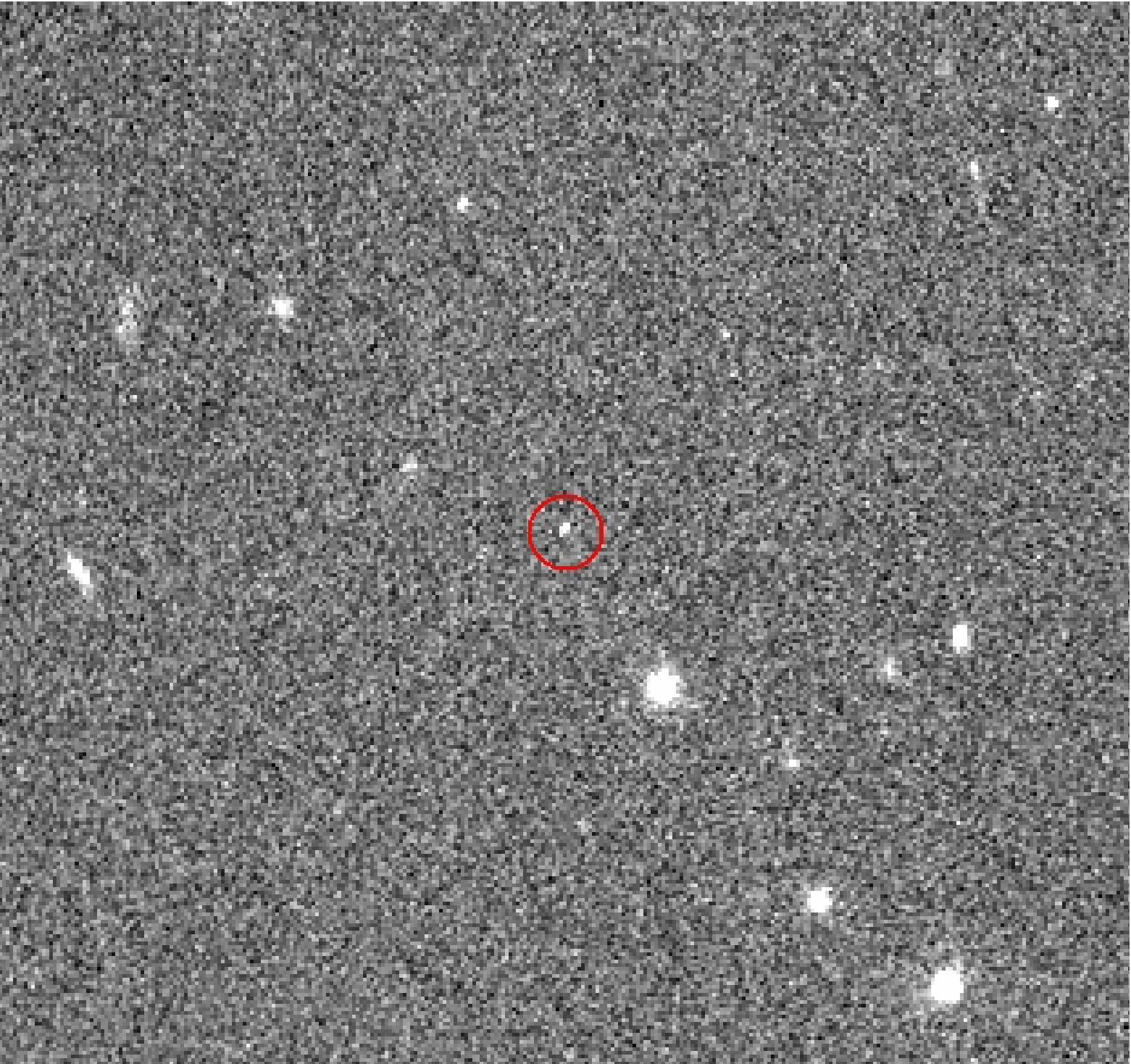}&
 \includegraphics[scale=0.4]{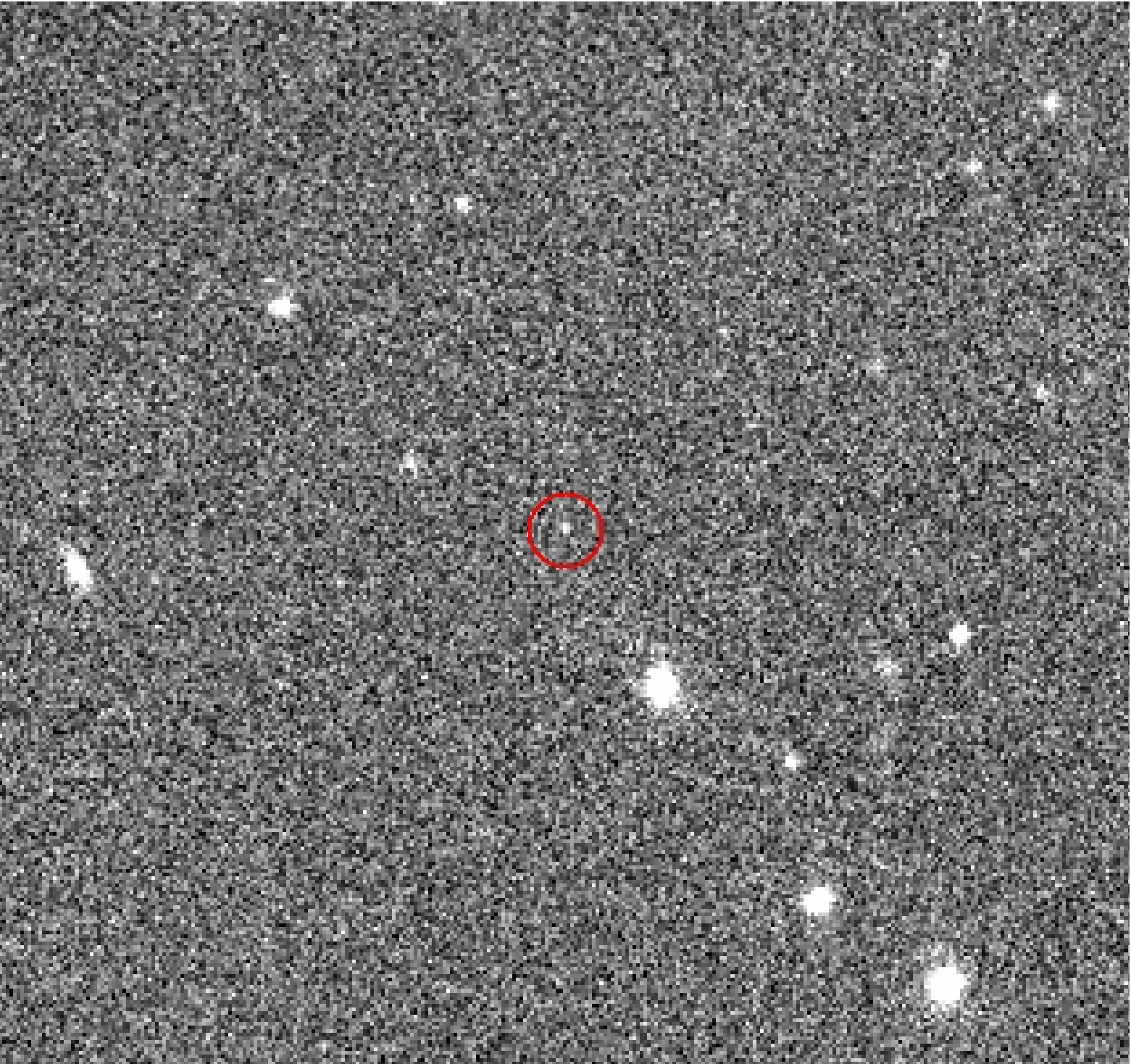}\\ \hline
\end{tabular}
 \caption{90\arcsec ~x 90\arcsec ~field centred on CFBDSIR1458. Upper
   left: $z'$ image from July 15, 2004; Upper right: $J$ image from
   March 6, 2009; Lower left: $H$ image from March 7, 2009; Lower
   right: $Ks$ image from March 7, 2009. The red circle
   highlights the position of CFBDSIR1458. North is up and  East is
   left.
\label{f_chart}}

\end{figure*}

CFBDSIR1458 colours are detailed in Table~\ref{colours} and  
shown in Fig.~\ref{f_chart}, and can be summarised as 
follows:
 \begin{itemize}
  \item Very red far-red colours: $z'-J=$3.94, which 
    Table~\ref{colours} compares to those of other
    ultracool brown dwarfs;
 \item Very blue NIR colours, with 
    [$J-H$;$J-K_{\rm s}$]=[-0.46;-0.94], pointing
    to very strong molecular absorptions in the $H$ and $K$ bands.
    The $J-K_{\rm s}$ colour of -0.94 approximately translates into a
    $J-K$ colour of -1.02 \citep{Stephens.2004,Leggett.2010}. 
 \end{itemize}

A second follow-up run at NTT on July 2009, confirmed 2 other
ultra-red objects, CFBDSIR221903.07+002417.92
and CFBDSIR221505.06+003053.11 (hereafter CFBDSIR2219 and
CFBDSIR2215), with even redder $z'-J$ colours, typical of ultracool
brown dwarfs. These objects have redder $z'-J$ than any known
T8+ brown dwarf. CFBDSIR2215, however, has only a moderately blue $J-H$
colour, which would be compatible with an earlier (i.e. T5-T7) spectral 
type, suggesting a peculiar spectrum. The
photometric properties  of these objects are described in Table~\ref{colours}
and Fig.~\ref{zJJH}.

\begin{table} 
\caption{(1) $J,~H,~Ks$ CFHT WIRCam Vega photometry and $z'_{AB}$ MegaCam 
  photometry of CFBDSIR1458. (2)$J,~H,~Ks$ NTT SOFI Vega photometry. 
 \label{mag}}
\begin{tabular}{|c|c|c|c|c|} \hline
Filter&$z'$&$J$&$H$&$K$s\\\hline
Mag(1)&$23.60\pm$0.24&19.66$\pm$0.02&20.12$\pm$0.13&20.60$\pm$0.37\\\hline
Mag(2)&-&19.72$\pm$0.04&19.96$\pm$0.16&20.13$\pm$0.32\\\hline
\end{tabular}
\end{table}

\begin{table} 
\caption{Colours of ultracool brown dwarfs.\label{colours}}
\begin{tabular}{|c|c|c|c|c|} \hline        
Object&$z'-J$&$J-H$&$J-K$\\\hline
ULAS0034$~^{1,3}$&3.90$\pm$0.06&-0.28$\pm$0.05&-0.12$\pm$0.06\\
CFBDS0059$~^{2}$&3.86$\pm$0.06&-0.21$\pm$0.06&-0.67$\pm$0.06\\
ULAS1335$~^{1,5}$&4.14$\pm$0.10&-0.35$\pm$0.01&-0.38$\pm$0.03\\
2MASS0939$~^{1,4}$&-&0.18$\pm$0.181&-0.58$\pm$0.10\\
WOLF940B$~^{6}$&3.99$\pm$0.13&-0.61$\pm$0.04&-0.69$\pm$0.05\\
CFBDSIR1458&3.94$\pm$0.24&-0.46$\pm$0.13&-1.02$\pm$0.37\\
CFBDSIR2219&4.45$\pm$0.10&-0.20$\pm$0.085&-\\
CFBDSIR2215&4.30$\pm$0.12&-0.11$\pm$0.08&-\\\hline
\end{tabular}
\begin{list}{}{}
  \item References:
 \item $^1$\citet{Leggett.2009}, $^2$\citet{Delorme.2008a}
  $^3$\citet{Warren.2007}, $^4$\citet{Tinney.2005},
  $^5$\citet{Burningham.2008}, $^6$\citet{Burningham.2009} 
\end{list}
\end{table}

 While CFBDSIR2219 and CFBDSIR2215 are queued for NIR spectroscopic 
 characterisation, we already obtained a spectrum of CFBDSIR1458,
 which we discuss in section 4.2.

{\bf Proper motion of CFBDSIR1458}\\
Since the $z'$ MegaCam image and the $J$ WIRCam image are co-aligned
by the CFBDSIR pipeline the proper motion is easily measured from 
these 2 images. CFBDSIR1458 moves by +0.58\arcsec ~in RA and 
-0.95\arcsec ~in DEC between the July 15, 2004 date of the $z'$
image and April 1, 2007 date of the $J$ image.
The centroiding and image alignment uncertainties are low, at
0.035\arcsec, but the error budget has to include chromatic
refraction and the uncertain parallactic motion. The estimated spectroscopic 
distance is $\sim$23pc, with a likely range of 15--30~pc 
(Table~\ref{pm}), which translates to a 0.04\arcsec maximum error 
from the parallax. Due to the very steep spectral energy distribution 
in the $z'$ band, chromatic refraction here is of the order of the 
measurement error. We also measured the proper motion between the 
$J$-band WIRCam discovery image and the $J$ band NTT follow-up 
image, finding +0.32\arcsec ~in RA and -0.80\arcsec ~in DEC between 
April 1, 2007 and March 5, 2009. This second measurement has
smaller centroiding and image-alignment uncertainties, 0.03\arcsec,
minimal chromatic refraction uncertainties since the two $J$-band observations
were obtained at similar airmasses, and a small unceretainty from
the parallax because the two observations were coincidentally obtained
at closely matched times of the year. It is therefore our preferred proper 
motion measurement, and is consistent with the previous one. 
Table~\ref{pm} presents the resulting yearly motion and kinematic
parameters.
  
\begin{table} 
\caption{Yearly proper motion for CFBDSIR1458. \label{pm}}
\begin{tabular}{|c|c|c|c|} \hline        
RA(\arcsec.yr$^{-1}$)       &   DEC(\arcsec.yr$^{-1}$)    & RA(km.s$^{-1}$)      &
DEC(km.s$^{-1}$)        \\ \hline
+0.17$\pm$0.016 & -0.41$\pm$0.016 &18  & 46 \\ \hline
\end{tabular}
\begin{list}{}{}
\item   The  systematic errors induced   by  chromatic
refraction and parallax effects are not  
  corrected. The estimate of the absolute tangential velocity assumes a
   distance of 23pc.
\end{list}
\end{table}

\subsection{Spectroscopic confirmation of a new ultracool brown dwarf}  
{\bf Observations and reduction}

 After reduction and analysis of the March 7, data, an ESO Director
 Discretionary Time observation request was  submitted on
 March 20, 2009 to obtain low-resolution NIR spectroscopy of CFBDSIR1458.
 This proposal for 4 hours of $H$-band observation, totalling 150
 minutes of exposure on 
 target, was accepted on April 22. The first
 observations were acquired on May 4, and the last on September 1,
 on average at high airmass, varying from 1.4 to more than 2.0. 

 The spectrum was extracted
and calibrated using our own IDL procedures.  The reduction proceeded
as follows.  The sequence of spectral images
were flat-fielded using an internal flat taken immediately after the
science frames. Since the trace was too faint for its position to be accurately
determined, its 
curvature was derived from the reference star spectrum. The frames were
then pair-subtracted, effectively removing most of the 
sky, dark current and hot pixels contributions. Each frame was
collapsed along the spectral dimension to determine the positive and
negative traces positions. We then extracted the spectra using positive
and negative extraction boxes that have identical but opposite
integrals; this minimised the contribution from residual sky line that
would have remained from the pair subtraction. The same operation was
performed 
on the A0 telluric calibration star.  Spectra derived from individual
image pairs were then median-combined into final target and calibration
star spectra. A telluric absorption spectrum was derived using the
calibration-star spectra. A black body spectrum with a temperature of 10
000K was assumed for the A0 star and hydrogen-lines were interpolated
over. 
The target spectrum was then divided by the derived telluric transmission
spectrum. A first-order wavelength calibration was obtained from an
argon-lamp 
spectrum, and fine-tuned by registering bright OH lines obtained from 
a sum of the pair of images of interest.\\

{\bf Spectroscopic properties}

\begin{figure}
\includegraphics[width=8cm]{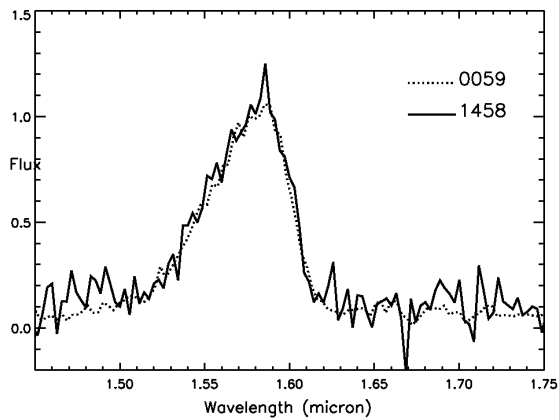}
\caption{$H$-band spectra of CFBSIR1458 compared with CFBDS0059, one
  the 2 coolest brown dwarfs known. Both spectra have been
  median-binned  over 6 pixels to match the full ISAAC spectral resolution of 500. 
\label{spectra}}
\end{figure}

\begin{figure}
\includegraphics[width=8cm]{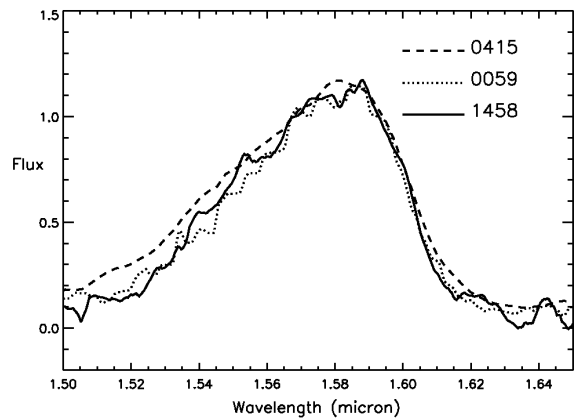}
\caption{Zoom on the $H$-band peak of the spectra of CFBSIR1458
  compared with CFBDS0059 and 
  2MASS0415 \citep{Burgasser.2003}, the T8 spectral template. The CFBSIR1458 ISAAC spectrum has
  been median-binned to a resolution of 170. 
\label{spectra_zoom}}
\end{figure}

The resulting spectrum (Fig.\ref{spectra}) has a low signal-to-noise
owing to the faintness of the target ($H_{vega}=20.12$) and its relatively high
airmass at the time of the observations. We plan to obtain better
signal-to-noise observations, as well as $J$ and $K$-band
low-resolution spectrum, but we were already able to derive from the present 
$H$-band  
spectrum the spectroscopic indices described by
\citet{Burgasser.2006} and \citet{Delorme.2008a}, which quantify the
strength of key molecular absorption bands. As shown in Table
\ref{indices}, this classifies CFBDSIR1458
as an ultracool brown dwarf with spectral type later than T8 and a
temperature in the same range ($\sim $500-600K) as the coolest brown
dwarfs known.  Direct comparison of the CFBDSIR1458 $H$-band spectrum with
other ultracool brown dwarfs (See Figs. \ref{spectra} and
\ref{spectra_zoom}) visually confirms that H$_2$O  and CH$_4$ 
absorption in its atmosphere are significantly stronger than they are on
2MASS0415 \citep{Burgasser.2003} the T8 spectral template. This
also strongly suggests that CFBDSIR1458 is indeed a later-than-T8
ultracool brown dwarf. The comparison with even cooler objects is less
clear cut, mainly because of the low signal-to-noise of the spectrum,
as emphasised by the strong variations in the spectral indices for 2
different -but both sufficiently sampled- binning of CFBDSIR1458
spectrum visible in Table \ref{indices}. However, both the spectrum and
the indices would tend to show that CFBDSIR1458 is not cooler than the
coolest brown dwarfs already known. 

This intermediate spectral feature would tentatively put CFBDSIR1458
in the same class as WOLF940B, that was classified as T8.5 by
\citet{Burningham.2009} who assigned a temperature of
550-600K to this object. Given the similarities in the $H$-band indices of
both objects, a reasonable estimate would put CFBDSIR1458 in the same
temperature range. 
However, this rough estimate will need to be 
confirmed by higher signal-to-noise multi-bands spectroscopic
observations. In particular, its very blue $J-K$ colour could be due to
sub-solar metallicity \citep{Leggett.2010}. Additional
observations would also 
enable looking for the NH$_3$ absorption in the CFBDSIR1458 spectrum,
such as the probable absorption band identified by
\citet{Delorme.2008a}.  IRAC imaging in the 4.5micron channel
  would also be very valuable since temperature can be reliably
  derived from the [$H-4.5$] colour \citep{Leggett.2010}.

\begin{table*} 
\caption{Spectral indices  of ultracool brown dwarfs.\label{indices}}     
\begin{tabular}{|c|c|c|c|c|} \hline   
Object       &Sp. Type&   H$_2$O-H  &      CH$_4$-H   &      NH$_3$-H\\ \hline
Gl570D& T7.5  &   0.208 &     0.137    &   0.672\\ \hline
2MASS0415& T8  &   0.183 &     0.104    &   0.625\\ \hline
WOLF940B&T8+/Y?   &   0.141  &    0.091  &     0.537\\ \hline
CFBDSIR1458 (1)&T8+/Y?   &   0.149  &    0.046  &     0.568\\ \hline
CFBDSIR1458 (2)&T8+/Y?   &   0.146  &    0.087  &     0.582\\ \hline
ULAS0034 & T8+/Y?  &  0.133    &  0.096   &    0.516\\ \hline
CFBDS0059& T8+/Y?   &  0.119   &   0.084    &   0.526\\ \hline
ULAS1335& T8+/Y?   &  0.114  &   0.077    &   0.564\\ \hline
\end{tabular}
\begin{list}{}{}
\item The 2 values for CFBDSIR1458 were
   derived 
  using a median-binning of the spectra over (1)  17 pixels (resolution
  $\sim$170) (2) 6 pixels(resolution 500). The other values are from \citet{Burgasser.2006,Warren.2007,Burningham.2008,Delorme.2008a,Burningham.2009} 
\end{list}
\end{table*}

\section{Conclusion}

  We have described CFBDSIR, a new NIR survey dedicated at finding
  ultracool brown dwarfs and using WIRCam camera on the the
  CFHT. Complementing existing deep far-red data by 
  new $J$-band observations, we select brown dwarfs
  candidates on their very red $z'-J$ colour. A robust PSF analysis
  allows us to derive reliable colours and to distinguish
  point-source-like brown dwarfs from most contaminants. The
  candidates are then confirmed by follow-up pointed NIR observations
  in $J$-band and confirmed ultracool brown dwarfs are imaged in
  $H$ and $K_{\rm s}$ bands. We used these photometric measurements to
  identify several new brown dwarfs, including 3 objects likely
  as cool as and possibly even cooler than any published brown dwarfs. 

 We  presented CFBSIR1458, the first CFBDSIR ultracool brown
 dwarf confirmed by spectroscopy. The analysis of its $H$-band spectra,
 though at relatively low-signal-to-noise, robustly confirms it as
 later than T8 spectral type and hints at a temperature in the
 550-600K range, so among the coolest brown dwarfs discovered. When the
 335 square degree survey is completed, we expect to discover a sample
 of 10 to 15  ultracool brown dwarfs, more than doubling the currently
 known 
 population of later than T8 objects and enabling study of them as a
 population rather than extreme individual objects. This will put
 strong constraints on cool stellar and planetary atmosphere, and with
 additional mid-infrared follow-up, will help to define the
 selection criteria for the upcoming WISE survey.

\begin{acknowledgements}
Thanks go to the queue
observers at CFHT who obtained data for this
paper. 
This research has made use of the VizieR catalogue access tool,
 of SIMBAD database and of Aladin, operated at CDS, Strasbourg. 
This research has benefitted from the M, L, and T-dwarf compendium
housed at DwarfArchives.org and maintained by Chris Gelino, Davy
Kirkpatrick, and Adam Burgasser.
  Financial support from the ''Programme National de Physique Stellaire" (PNPS)
of CNRS/INSU, France, is gratefully acknowledged.
\end{acknowledgements}

\bibliographystyle{aa}
\bibliography{bib}

\end{document}